\documentclass[12pt]{article}

\usepackage{amssymb,amsmath,amsfonts,amssymb}
\usepackage{graphics,graphicx,color,hhline,euscript}
   
\def\@abssec#1{\vspace{.05in}\footnotesize \parindent .2in
{\bf #1. }\ignorespaces}

\graphicspath{{/EPSF/}{Figures/}}
 
\newtheorem{theorem}{Theorem}[section]
\newtheorem{lemma}[theorem]{Lemma}

\newtheorem{corollary}[theorem]{Corollary}

\def \Rm {\mathbb R}

\newcommand{\E}{\mathbb E}
\newcommand{\bbP}{\mathbb P}

\newcommand{\dsum}{\displaystyle\sum}
\newcommand{\dint}{\displaystyle\int}
\newcommand{\pdr}[2]{\dfrac{\partial{#1}}{\partial{#2}}}
\newcommand{\pdrr}[2]{\dfrac{\partial^2{#1}}{\partial{#2}^2}}

\newcommand{\bk}{\mathbf k}

\newcommand{\bp}{\mathbf p} \newcommand{\bq}{\mathbf q}
\newcommand{\bu}{\mathbf u} \newcommand{\bv}{\mathbf v}
\newcommand{\bw}{\mathbf w}
\newcommand{\bx}{\mathbf x} \newcommand{\by}{\mathbf y}
\newcommand{\bz}{\mathbf z}

\newcommand{\calQ}{\mathcal Q}
\newcommand{\calL}{\mathcal L}

\newcommand{\calG}{\mathcal G} 
\newcommand{\calK}{\mathcal K} 
\newcommand{\calF}{\mathcal F}

\newcommand{\calC}{\mathcal C}

\newcommand{\bzero}{\mathbf 0}

\newcommand{\bxi}{\boldsymbol \xi} 
\newcommand{\bzeta}{\boldsymbol \zeta}

\newcommand{\cout}[1]{}

\newcommand{\mG}{\mathcal G}

\def\un{{\mathbf{1}}\hspace{-0.24em}\mathrm{I}}

 \renewcommand{\arraystretch}{1.5}

\title{Self-averaging of kinetic models for waves in random media}

\author{Guillaume Bal \thanks{Department of Applied Physics and
    Applied Mathematics, Columbia University, New York NY, 10027;
    gb2030@columbia.edu; 
   } \and Olivier Pinaud
  \thanks{Universit\'e de Lyon,
Universit\'e Lyon 1, CNRS, UMR 5208 Institut Camille Jordan/ISTIL,
B\^atiment du Doyen Jean Braconnier,
43, blvd du 11 novembre 1918,
F - 69622 Villeurbanne Cedex,
France;   
  pinaud@math.univ-lyon1.fr} }

\begin{document}
 
\maketitle


\begin{abstract}
  Kinetic equations are often appropriate to model the energy density
  of high frequency waves propagating in highly heterogeneous media.
  The limitations of the kinetic model are quantified by the
  statistical instability of the wave energy density, i.e., by its
  sensitivity to changes in the realization of the underlying
  heterogeneous medium modeled as a random medium. In the simplified
  It\^o-Schr\"odinger regime of wave propagation, we obtain optimal
  estimates for the statistical instability of the wave energy density
  for different configurations of the source terms and the domains
  over which the energy density is measured.  We show that the energy
  density is asymptotically statistically stable (self-averaging) in
  many configurations.  In the case of highly localized source terms,
  we obtain an explicit asymptotic expression for the scintillation
  function in the high frequency limit.
\end{abstract}
 

\noindent{\bf keywords:} Waves in random media, kinetic models, 
It\^o-Schr\"odinger equation, statistical instability

\renewcommand{\thefootnote}{\fnsymbol{footnote}}
\renewcommand{\thefootnote}{\arabic{footnote}}

\renewcommand{\arraystretch}{1.1}


\section{Introduction}
\label{sec:intro}

Let us consider the following scalar wave equation for the pressure
potential $p(\tau,\bx,t)$:
\begin{equation}
  \label{eq:wave}
  \dfrac{1}{c^2(\bx,t)} \pdrr p\tau = \Delta_\bx p +
  \pdrr pt,
\end{equation}
where $\tau$ is time, $(\bx,t)\in\Rm^d\times\Rm$ denote the spatial
variables, $\Delta_\bx$ is the Laplace operator in the transverse
variables $\bx$, and $c(\bx,t)$ is the local sound speed. Our
objective is to understand the properties of $p(\tau,\bx,t)$ when
$c(\bx,t)$ is a highly oscillatory random field and the initial
conditions for $p(\tau,\bx,t)$ oscillate at the same frequency.

The analysis of high frequency waves in random media based on
\eqref{eq:wave} is extremely complicated and still not totally
established mathematically. Since the wave field is oscillatory, its
(weak) limit typically misses most of the energy of the wave field
$p$. Kinetic models are then used to capture the energy density of the
wave fields; see e.g.  \cite{BKR-liouv,BPR-SD02,Erdos-Yau2,LS-ARMA-07}
for rigorous results, \cite{B-WM-05,RPK-WM} for more formal
derivations, and \cite{chandra,ishimaru,LN-EPJB99,sheng} for
references in the physical literature.

The validity of the kinetic model is limited by its statistical
instability, namely by its variability when the realization of the
underlying random medium is changed. In many situations, the energy
density is self-averaging \cite{B-Ito-04,BKR-liouv,BPR-SD02}, which
means that the energy density measured (averaged) on a (sufficiently
large) domain is asymptotically, as the frequency goes to infinity,
independent of the realization of the random medium. The above results
often require that the domain of measurements be of size independent
of the wavelength and that the source term for the kinetic model be
sufficiently smooth.

In this paper, we are interested in the statistical stability of such
kinetic models in a very simplified regime of wave propagation, namely
the It\^o-Schr\"odinger regime. The latter regime arises when the wave
field is a very narrow beam propagating in the direction $t$ and the
sound speed $c(\bx,t)$ oscillates more rapidly in the direction $t$
than it does in other directions. Such assumptions are valid in
somewhat restrictive practical settings.  However, this regime of wave
propagation is relatively simple to analyze mathematically and
provides interesting qualitative answers regarding the statistical
stability of more general kinetic models.

The validity of kinetic models has been analyzed numerically in
several settings \cite{BP-WM06,BP-MMS-07,BRe-SIAP-08}, with quite good
agreements with the energy density given by wave equations of the form
\eqref{eq:wave}. Such kinetic models may then be used to solve inverse
problems, where constitutive parameters in the transport equation
modeling e.g. buried inclusions or statistics of the random medium,
are reconstructed from available boundary measurements.  We refer the
reader to \cite{BP-MMS-07,BRe-SIAP-08} for reconstructions based on
synthetic (numerical) data and to \cite{BCLR-IP-07} for kinetic
reconstructions from experimental data in the micro-wave regime; see
also \cite{B-IAP07-08} for a review on the use of kinetic models in
the imaging of buried inclusions.  These studies show that the kinetic
models perform relatively well. Their limitations are almost entirely
caused by our lack of knowledge of the random medium, which generates
some statistical instabilities in the measurements. Understanding
these instabilities will allow us to improve on the reconstructions
and to have a better understanding of the maximal resolution that can
be achieved.

\medskip

\noindent{\bf It\^o-Schr\"odinger regime.}
In the It\^o-Schr\"odinger regime, we introduce $\psi(\bx,t;\kappa)$
as
\begin{equation}
  \label{eq:psi}
  p(\tau,\bx,t)=\dfrac{1}{2\pi}\dint_\Rm e^{i\kappa(t-c_0\tau)}
   \psi(\bx,t;\kappa) c_0 d\kappa,
\end{equation}
where $c_0$ is the background sound speed, assumed to be constant.
Thus $\psi$ represents waves at position $(\bx,t)$ propagating with
frequency $\omega=c_0|\kappa|$. After appropriate scalings and
simplifications, the wave field $\psi$ satisfies the following
It\^o-Schr\"odinger stochastic partial differential equation:
\begin{equation}
  \label{eq:itosch}
  d\psi_\eta(\bx,t;\kappa)=\dfrac12 
    \big(i\eta \Delta_\bx - \kappa^2 R(\bzero)\big)
    \psi_\eta dt + i\kappa \psi_\eta B\Big(\dfrac{\bx}\eta,dt\Big).
\end{equation}
Since $\kappa$ plays no significant role in the sequel, we set it to
$\kappa=1$. Here, $B(\bx,dt)$ is the standard Wiener measure, whose
statistics are described by
\begin{equation}
  \label{eq:B}
  \E\{B(\bx,t)B(\by,t')\} = R(\bx-\by) t \wedge t',
\end{equation}
where $\E$ is mathematical expectation with respect to the measure of
an abstract probability space on which $B(\bx,dt)$ is defined and
$t\wedge t'=\min(t,t')$.  We
shall not justify \eqref{eq:itosch} from \eqref{eq:wave}. See
\cite{BCF-SIAP-96} for a justification in one dimension of space and
\cite{B-Ito-04} for the scaling arguments leading to
\eqref{eq:itosch}.

For our purposes, $\psi_\eta(\bx,t)$ satisfies a wave equation with
highly oscillatory coefficients oscillating at a frequency inversely
proportional to the small parameter $\eta\ll1$. We assume that
$\psi_\eta(\bx,0)$ also oscillates at a frequency comparable to
$\eta^{-1}$ and are interested in the properties of the wave field as
$\eta\to0$. Because the field oscillates rapidly, its weak limit is of
little interest. A more interesting quantity is the energy density of
the waves $|\psi_\eta|^2(\bx,t)$, or the probability density in the
context of quantum waves. Because the energy density does not satisfy
a closed-form equation, it is more convenient to analyze energy
densities by introducing the following Wigner transform of the wave
field:
\begin{equation}
  \label{eq:Wigner}
  W_\eta(t,\bx,\bk) = \dfrac{1}{(2\pi)^d}\dint_{\Rm^d}
   e^{i\bk\cdot\by} \psi_\eta\Big(\bx-\dfrac{\eta\by}2,t\Big)
    \overline{\psi_\eta}\Big(\bx+\dfrac{\eta\by}2,t\Big) d\by,
\end{equation}
where $\overline{\psi_\eta}$ denotes complex conjugation of $\psi$.
Note that $\int_{\Rm^d} W_\eta(t,\bx,\bk)d\bk=|\psi_\eta(\bx,t)|^2$ by
inverse Fourier transform so that $W_\eta$ may be seen as a phase
space (microlocal) decomposition of the energy density.

Let $\psi_\eta(\bx,0)$ be a sequence of functions uniformly bounded in
$L^2(\Rm^d)$, $\eta$-oscillatory, and compact at infinity in the sense
of \cite{GMMP}, i.e., such that for every continuous compactly
supported function $\varphi$ on $\Rm^d$, we have:
\begin{displaymath}
  \begin{array}{l}
    \overline{\lim\limits_{\eta\to0}} \dint_{|\bk|>R/\eta}
   |\widehat{\varphi \psi_\eta}(\bk)|^2d\bk \to0,\quad
   \mbox{ as } R\to\infty \\
   \overline{\lim\limits_{\eta\to0}}  \dint_{|\bx|>R}
    |\psi_\eta|^2(\bx)d\bx \to0,\quad
   \mbox{ as } R\to\infty.
  \end{array}
\end{displaymath}
A practical sufficient condition is that $\psi_\eta(\bx,0)$ is
compactly supported and $\eta\nabla\psi_\eta(\bx,0)$ is square
integrable with $L^2(\Rm^d)$-norm bounded independently of $\eta$.
Then, we have the following convergence result \cite{GMMP,LP}: The
Wigner transform $W_\eta(0,\bx,\bk)$ converges, after possible
extraction of subsequences, in the space of distributions $\mathcal
D'(\Rm^{2d})$ to a Radon measure $W_0(0,\bx,\bk)$, and moreover, we
have
\begin{equation}
  \label{eq:consenerg}
  \dint_{\Rm^{2d}} W_0(0,\bx,\bk) d\bx d\bk = \lim\limits_{\eta\to0}
   \dint_{\Rm^d} |\psi_\eta|^2(\bx,0) d\bx.
\end{equation}
In other words, the limiting Wigner transform captures all the energy
of the incident wave field $\psi_\eta$ in the limit $\eta\to0$.

\medskip

\noindent{\bf Kinetic Model.}
Upon using the It\^o formula, we obtain that the average Wigner
transform
\begin{equation}
  \label{eq:aeta}
  a_\eta(t,\bx,\bk) = \E\{W_\eta(t,\bx,\bk)\},
\end{equation}
solves the following kinetic equation
\begin{equation}
  \label{eq:kinetic}
  \begin{array}{l}
  \pdr {a_\eta}t + \bk\cdot\nabla_\bx a_\eta + R_0a_\eta
  = \dint_{\Rm^d} \hat R(\bk-\bq)  a_\eta(t,\bx,\bq)\dfrac{d\bq}{(2\pi)^d},\\
   a_\eta(0,\bx,\bk) = W_\eta(0,\bx,\bk),
  \end{array}
\end{equation}
where we assume that $\psi_\eta(\bx,0)$, whence $a_\eta(0,\bx,\bk)$,
is deterministic; see e.g. \cite{B-Ito-04} for the details of the
derivation.  We have defined $R_0=R(\bzero)$ and $\hat R(\bk)$ as the
Fourier transform of $R(\bx)$, with the convention that
\begin{equation}
  \label{eq:FT}
  \hat R(\bk) = \mathcal F R (\bk) = 
  \dint_{\Rm^d} e^{-i\bk\cdot\bx} R(\bx) d\bx.
\end{equation}
Since $R(\bx)$ is a correlation function, $\hat R(\bk)$ is
non-negative by Bochner's theorem.  For the rest of the paper, we
assume that $\hat R(\bk)\in L^1(\Rm^d)\cap L^\infty(\Rm^d)$.  Note
that $\int_{\Rm^{2d}} a_\eta(t,\bx,\bk)d\bx d\bk$ is independent of
time so that the total energy of the initial condition is preserved by
the transport evolution.

\medskip

\noindent{\bf Scintillation.}
The validity of the kinetic model \eqref{eq:kinetic} to describe the
ensemble averaging of the phase space energy density of the wave field
is trivial in the It\^o-Schr\"odinger regime: the kinetic model
\eqref{eq:kinetic} is here exact for all $\eta\geq0$, unlike what
happens in other regimes of wave propagation
\cite{BKR-liouv,BPR-SD02,RPK-WM}. It remains however to understand how
stable it is. In other words, how good an approximation is
$a_\eta(t,\bx,\bk)$ of the random field $W_\eta(t,\bx,\bk)$. A natural
object in the study of the statistical stability of $W_\eta$ is the
following covariance function:
\begin{equation}
  \label{eq:Jeta}
  J_\eta(t,\bx,\bk,\by,\bp) = \E\{W_\eta(t,\bx,\bk)W_\eta(t,\by,\bp)\}
   - \E\{W_\eta(t,\bx,\bk)\}\E\{W_\eta(t,\by,\bp)\}.
\end{equation}
We refer to this function as the scintillation function, in analogy to
how stars are perceived to twinkle because the realization of the
atmosphere changes in time.

We shall see that the size of the scintillation function crucially
depends on the smoothness of the initial conditions $\psi_\eta(\bx,0)$
and $a_\eta(0,\bx,\bk)$ and on the support of the domain over which
the energy density is averaged. The effect of the averaging will be
quantified by measuring $J_\eta$ in appropriate (weak) norms.

One of the main advantages of the It\^o-Schr\"odinger regime of wave
propagation is that $J_\eta(t,\bx,\bk,\by,\bp)$ satisfies a closed
form equation. Another application of the It\^o formula \cite{B-Ito-04}
shows that $J_\eta$ is the solution of the following kinetic equation:
\begin{equation}
  \label{eq:kinfour}
  \Big(\pdr{}t + \mathcal T_2 + 2R_0 - \mathcal Q_2 - \mathcal K_\eta\Big)
    J_\eta = \mathcal K_\eta a_\eta \otimes a_\eta,
\end{equation}
with vanishing initial conditions $J_\eta(0,\bx,\bk,\by,\bp)=0$, where
\begin{equation}
  \label{eq:ops}
  \begin{array}{rcl}
    \mathcal T_2 &=& \bk\cdot\nabla_\bx + \bp\cdot\nabla_\by \\
    \mathcal Q_2 J &=& \dint_{\Rm^{2d}}
   \Big(\hat R(\bk-\bk')\delta(\bp-\bp')+\hat R(\bp-\bp')\delta(\bk-\bk')
    \Big) J(\bk',\bp') \dfrac{d\bk' d\bp'}{(2\pi)^d} \\
    \mathcal K_\eta h &=& \dsum_{\epsilon_i,\epsilon_j=\pm1}
      \dint_{\Rm^{2d}}\hat R(\bu) e^{i\frac{(\bx-\by)\cdot\bu}{\eta}}
      \epsilon_i\epsilon_j h(\bx,\bk+\epsilon_i\dfrac\bu2,
     \by,\bp+\epsilon_j\dfrac\bu2) \dfrac{d\bu}{(2\pi)^d}.
  \end{array}
\end{equation}

In the absence of the operator $\mathcal K_\eta$, the variables
$(\bx,\bk)$ and $(\by,\bp)$ remain uncoupled in \eqref{eq:kinfour} and
the scintillation vanishes. Scintillation is created as the waves
propagate through the random medium with a rate of creation
proportional to $K_\eta a_\eta\otimes a_\eta$. Notice that $K_\eta$
involves a highly oscillatory integral. Outside of the diagonal
$\bx=\by$, this oscillatory integral is small, whereas in the vicinity
of the diagonal $\bx=\by$, it is not. We thus observe that $K_\eta h$
is small when $h$ is smooth and large when part of $h$ is concentrated
near $\bx=\by$.

\medskip

\noindent{\bf Outline.} The rest of the paper is structured as follows.
The main results of the paper are summarized in section
\ref{sec:main}. We obtain estimates for $J_\eta$ in various norms, and
in the specific case of initial conditions for $a_\eta$ of the form
$a_\eta(0,\bx,\bk)=\delta(\bx)f(\bk)$, show that $\eta^{-1}J_\eta$
converges to a measure $J$ solving an explicit kinetic equation.
Section \ref{sec:stab} presents stability estimates for the
scintillation operator $K_\eta$ defined in \eqref{eq:ops} and for the
kinetic equations \eqref{eq:kinetic} and \eqref{eq:kinfour}.  The
proof of the stability estimates for $J_\eta$ are given in section
\ref{sec:estimates} whereas the proof of convergence of
$\eta^{-1}J_\eta$ when $a_\eta(0,\bx,\bk)=\delta(\bx)f(\bk)$ is given
in section \ref{sec:conv}.

\section{Main results}
\label{sec:main}

Let $\psi_\eta(\bx,0)$ be a sequence of $\eta-$oscillatory, compact at
infinity, functions uniformly bounded in $L^2(\Rm^d)$. This is the
case of interest for us here, where we can define the Wigner transform
\eqref{eq:Wigner} and pass to the high frequency limit $\eta\to0$ while
still ensuring that energy is conserved as in \eqref{eq:consenerg}.
We are interested in quantifying the statistical stability of the
Wigner transform $W_\eta(t,\bx,\bk)$ and do so by analyzing the
scintillation function $J_\eta$ defined in \eqref{eq:Jeta}.

We present two results. The first result proposes an upper bound for
$J_\eta$ in different norms and for different initial conditions
$\psi_\eta(\bx,0)$. The second result analyzes the convergence
properties of $J_\eta$ as $\eta\to0$ for initial conditions of the
form $a_\eta(0,\bx,\bk)=\delta(\bx)f(\bk)$, which correspond to
localized sources at position $\bx=\bzero$ radiating energy smoothly
in wavenumber $\bk$. In this context, we will show that $J_\eta$ is of
order $O(\eta)$ and will obtain the limit of $\eta^{-1}J_\eta$ as
$\eta\to0$.

\medskip

\noindent{\bf Some typical initial conditions.}
Let us consider initial conditions $\psi_\eta(\bx,0)$ oscillating at
frequencies of order $\eta^{-1}$ and with a spatial support of size
$\eta^{\alpha}$ for $0\leq\alpha\leq 1$. The parameter $\alpha$
quantifies the macroscopic concentration of the initial condition.

The simplest example is a modulated plane wave of the form:
\begin{equation}
  \label{eq:planewave}
  \psi_\eta^{(1)} (\bx) = \dfrac{1}{\eta^{\frac{d\alpha}2}}
   \chi\Big( \dfrac{\bx}{\eta^\alpha}\Big) 
   e^{i\frac{\bx\cdot\bk_0}\eta},
\end{equation}
where $\chi(\bx)$ is a smooth compactly supported function on $\Rm^d$.
The direction of propagation is given by $\bk_0$. Note that the above
sequence of initial conditions is indeed uniformly bounded in
$L^2(\Rm^d)$, compact at infinity, and $\eta$-oscillatory.

As another example of initial conditions, we consider
\begin{equation}
  \label{eq:sphericalwave}
  \psi_\eta^{(2)} (\bx) = \dfrac{1}{\eta^{\frac{(d-1)\alpha+1}2}}
   \chi\Big( \dfrac{\bx}{\eta^\alpha}\Big) 
   J_0\Big(\dfrac{|\bk_0||\bx|}\eta\Big),
\end{equation}
where $J_0$ is the zero-th order Bessel function of the first kind.
Such an initial condition is supported in the Fourier domain in the
vicinity of wavenumbers $\bk$ such that $|\bk|=|\bk_0|$ so that
$\psi_\eta^{(2)}$ emits radiation isotropically at wavenumber
$|\bk_0|$; see \cite{BP-WM06,BP-MMS-07} for more details. We again
verify that the above sequence of initial conditions is indeed
uniformly bounded in $L^2(\Rm^d)$, compact at infinity, and
$\eta$-oscillatory. For this, we use that $J_0(z)=\sqrt{\frac{2}{\pi
    z}}\cos(z-\frac\pi4)+O(z^{-3/2})$.

\medskip

\noindent{\bf Domain of measurements.} For the above initial conditions
for $\psi_\eta$, we are interested in the corresponding Wigner
transform $W_\eta(t,\bx,\bk)$ and scintillation function $J_\eta$. It
turns out that $J_\eta$ is itself oscillatory so that its size depends
on the scale at which it is measured. In order to capture this scale,
we introduce a test function $\varphi\in \mathcal S(\Rm^{2d})$, a
fixed wavenumber $\bk_1\in\Rm^d$, and define
\begin{equation}
  \label{eq:varphi12}
  \varphi_{\eta,s_1,s_2}(\bx,\bk) = \dfrac{1}{\eta^{d(s_1+s_2)}}
   \varphi\Big(\dfrac{\bx}{\eta^{s_1}},\dfrac{\bk-\bk_1}{\eta^{s_2}}\Big).
\end{equation}
We then denote by $\langle \cdot,\cdot\rangle$ the duality product
$\mathcal S'(\Rm^{n})$-$\mathcal S(\Rm^{n})$ for $n=2d$ or $n=4d$ and
want to quantify $\langle W_\eta, \varphi_{\eta,s_1,s_2} \rangle$, the
energy density averaged over a domain (in the phase space) of width
$\eta^{s_1}$ in space and $\eta^{s_2}$ in wavenumbers.

By using the Chebyshev inequality, we obtain the following estimate on
the probability that $W_\eta$ deviate from its ensemble average
$a_\eta$:
\begin{equation}
  \label{eq:Cheb}
  \bbP\Big(|\langle W_\eta(t),\varphi_{\eta,s_1,s_2}\rangle
    - \langle a_\eta(t),\varphi_{\eta,s_1,s_2}\rangle | \geq\delta \Big)
   \leq \dfrac{1}{\delta^2} 
    \langle J_\eta(t),\varphi_{\eta,s_1,s_2} \otimes \varphi_{\eta,s_1,s_2}
  \rangle.
\end{equation}
Here, $a\otimes a(\bx,\bk,\by,\bp)=a(\bx,\bk) a(\by,\bp)$.  In other
words, when the above right-hand side converges to $0$, then we find
that $\langle W_\eta(t),\varphi_{\eta,s_1,s_2}\rangle$ converges in
probability to $0$, which implies that $W_\eta(t)$ converges weakly
and in probability to $0$. The measured energy density is thus
asymptotically statistically stable. A very relevant practical
question pertains to the largest values of $s_1$ and $s_2$ that can be
chosen so that the Wigner transform is still statistically stable in
the limit $\eta\to0$. We are now ready to state our main theorem on
this issue.

\medskip

\noindent{\bf Bounds for the scintillation function.}
For any $\varphi(\bx,\bk)\in L^2(\Rm^{2d})$, let $\mathcal
F_\bx\varphi(\bu,\bk)$ and $\mathcal F_\bk\varphi(\bx,\bxi)$ be the
Fourier transforms of $\varphi$ in the first variable only and in the
second variable only, respectively.  We also denote by $a\lesssim b$
the inequality $a\leq Cb$, where $C>0$ is some universal constant.  Then
we have the following result:
\begin{theorem}
  \label{thm:1}
  Let $\psi_\eta(\bx,0)$ be a sequence of functions uniformly bounded
  in $L^2(\Rm^d)$, compact at infinity, and $\eta$-oscillatory. Let
  $a_\eta(0,\bx,\bk)$ be the corresponding sequence of Wigner
  transforms given by \eqref{eq:Wigner}. We assume that $\mathcal
  F_\bx a_\eta(0)$ and $\mathcal F_\bk a_\eta(0)$ are integrable
  functions and that
  \begin{equation}
    \label{eq:bdsL1}
    \|\mathcal F_\bx a_\eta(0,\bu,\bk)\|_{L^1(\Rm^{2d})}
   \lesssim \eta^{-\alpha d} \quad \mbox{ and } \quad
     \|\mathcal F_\bk a_\eta(0,\bx,\bxi)\|_{L^1(\Rm^{2d})}
   \lesssim \eta^{-\beta d},
  \end{equation}
  for some $\alpha\in\Rm$ and $\beta\in\Rm$. Then we find that 
  \begin{equation}
    \label{eq:bdJeta}
    \langle J_\eta(t),\varphi_{\eta,s_1,s_2} \otimes \varphi_{\eta,s_1,s_2}
  \rangle \lesssim \eta^{(\alpha-\beta)\vee0 -2ds_2} 
   \big(\eta^{d(1-\alpha-2s_1)} \wedge \eta^{d(1-2\alpha-s_1)}\big).
  \end{equation}
  Here, $a\wedge b=\min(a,b)$ and $a\vee b=\max(a,b)$.
\end{theorem}

Of interest here is the following corollary:
\begin{corollary}
  \label{cor:1}
  Let $\psi_\eta(0)$ be given by one of the expressions in
  \eqref{eq:planewave} or \eqref{eq:sphericalwave}. Then
  \eqref{eq:bdJeta} holds with $\beta=1-\alpha$.
\end{corollary}

We can deduce the following results from the above corollary. In what
follows, we consider that averaging takes place over a large domain of
wavenumbers so that $s_2=0$, as e.g., in spatial measurements of the
physical energy density. 

\medskip

\noindent{\bf Support of the sources.} 
Let us assume that the spatial support of the domain of measurements
is large so that $s_1=0$ as well. Then we find that
\begin{equation}
  \label{eq:Jeta1}
  \langle J_\eta(t),\varphi\otimes\varphi\rangle \lesssim 
   \eta^{\alpha+d(1-\alpha)}.
\end{equation}
In other words, the scintillation is of order $O(\eta^d)$ when
$\alpha=0$, which corresponds to a large support of the initial source
term.  This corresponds to the ideal case where the scintillation is
smallest. In such a setting, we obtain that $\langle
W_\eta-a_\eta,\varphi\rangle$ is of order $\eta^{\frac d2}$. This is the
most stable situation.

For a very narrow support of the initial source term comparable to the
correlation length of the medium, namely when $\alpha=1$, we obtain
that the scintillation is of order $O(\eta)$ so that $\langle
W_\eta-a_\eta,\varphi\rangle$ is now of order $\eta^{\frac12}$. We
thus obtain statistical stability of the energy density generated by a
very localized source term whose radiation pattern in $\bk$ is smooth,
although the statistical instability is much larger than in the case
$\alpha=0$.  We know that for sources that are highly localized both
in space and in wavenumbers, the scintillation does not converge to
$0$ and the energy density is not asymptotically statistically stable;
see \cite{B-Ito-04}. Such highly localized initial conditions would
correspond to a choice $\alpha=\beta=1$ in Theorem \ref{thm:1}.  We
will confirm in the next theorem that the order $O(\eta)$ above is
optimal.

\medskip

\noindent{\bf Small domain of measurements.} Conversely, we can consider the 
case of a source term with a large support, which corresponds to
$\alpha=0$, and a very small measurement domain. In this setting, we
find that
\begin{equation}
  \label{eq:Jeta2}
  \langle J_\eta,\varphi_{\eta,s_1}\otimes\varphi_{\eta,s_1}\rangle
   \lesssim \eta^{d(1-s_1)}.
\end{equation}
This means that the energy density becomes asymptotically
statistically stable as soon as it is measured over an area that is
large compared to the correlation length of the medium. This is an
optimal result of self-averaging as we cannot expect the energy
density to be statistically stable point-wise, or when averaged over
sub-wavelength domains. The above result, which is based on estimating
$K_\eta$ in \eqref{eq:ops} in appropriate norms, improves on estimates
obtained in \cite{B-Ito-04,PRS-MMS-07}.

We can also consider intermediate situations where both the source and
the measurement domain have small support. In that case, the optimal
estimate for the scintillation depends on whether $\alpha<s_1$ or
$s_1<\alpha$. These results are in fact optimal when the source term
and the domain of measurements are located at the same place. Such a
geometry explains why we do not obtain scintillation proportional to
$\eta^{d(1-\alpha-s_1)}$. We should obtain better estimates when the
domain of measurements and the source term are not centered around the
same point, though this cannot be inferred from our current results.

\medskip

\noindent{\bf Convergence of scintillation.}
Let us consider the case of initial conditions of the form
\eqref{eq:planewave} or \eqref{eq:sphericalwave} with $\alpha=1$,
i.e., for tightly localized source terms, in (transverse) dimension
$d\geq2$. The Wigner transform of such source terms converges in the
limit $\eta\to0$ to a distribution of the form $\delta(\bx)f(\bk)$,
where $f(\bk)$ is a smooth function when $\chi(\bx)$ is smooth
\cite{LP}. We consider the kinetic equations with such initial
conditions and obtain the following result.
\begin{theorem}
  \label{thm:2} Let $J_\eta$ be the solution of \eqref{eq:kinfour} with
  the initial condition in \eqref{eq:kinetic} given by
  $a_\eta(0,\bx,\bk)=\delta(\bx) f(\bk)$ for some smooth function
  $f(\bk)$ in dimension $d\geq2$. Then $\eta^{-1}J_\eta(t)$ converges
  in the space of distributions uniformly in time to the limit $J(t)$,
  which solves the following kinetic equation
  \begin{equation}
    \label{eq:J}
     \Big(\pdr{}t + \mathcal T_2 + 2R_0 - \mathcal Q_2 \Big)
    J = J^0,
  \end{equation}
  where 
  \begin{equation}
    \label{eq:J0}
    \begin{array}{ll}
    J^0(t) = &e^{-2R_0t}\delta(\bx-t\bp)\delta(\by-t\bq) 2\pi\times \\&
   \!\!\!\dint_{\Rm^d}
   \!\hat R(\bu) \delta\big(\bu\cdot(\bp-\bq)\big) 
    f(\bp-\dfrac{\bu}2)\Big( f(\bq-\dfrac{\bu}2)-f(\bq+\dfrac{\bu}2)\Big)d\bu.
   \end{array}
  \end{equation}
\end{theorem}
The above theorem should be interpreted as follows. As time
propagates, the transport ballistic part
$a^0(t,\bx,\bk)=e^{-R_0t}\delta(\bx-t\bk)f(\bk)$ creates some
instabilities, which converge after appropriate scaling to $J^0(t)$.
The scintillation thus generated is then transported by the
transport equation \eqref{eq:J}. We also observe that the error
estimate of order $O(\eta)$ in \eqref{eq:Jeta1} with $\alpha=1$ is
optimal.

\section{Functional setting and stability estimates}
\label{sec:stab}

In preparation for the proof of the theorems and the corollary
presented in the preceding section, we prove here some stability
results for the transport equations \eqref{eq:kinetic} and
\eqref{eq:kinfour} and for the scintillation operator $K_\eta$.

We denote by $\mathcal F$ the operator of Fourier transform with
respect to all variables of the function on which it applies. For
$1\leq p\leq \infty$, we introduce $X_p$ as the subspace of 
tempered distributions in $\mathcal S'(\Rm^{4d})$ such that
\begin{equation}
  \label{eq:Xp}
  \|h\|_{X_p}^p = \sup\limits_{\bv,\bzeta\in\Rm^d} \dint_{\Rm^d}
   \sup\limits_{\bxi\in\Rm^d} |\mathcal F h(\bu,\bxi,\bv,\bzeta)|^p
   d\bu <\infty,
\end{equation}
for $1\leq p<\infty$ and 
\begin{equation}
  \label{eq:Xinfty}
  \|h\|_{X_\infty} = \sup\limits_{\bu,\bzeta,\bv,\bxi\in\Rm^d}
    |\mathcal F h(\bu,\bxi,\bv,\bzeta)| <\infty.
\end{equation}
We also define $Y_p$  as the subspace of 
tempered distributions in $\mathcal S'(\Rm^{2d})$ such that
\begin{equation}
  \label{eq:Yp}
  \|g\|_{Y_p}^p = \dint_{\Rm^d}
   \sup\limits_{\bxi\in\Rm^d} |\mathcal F g(\bu,\bxi)|^p
   d\bu <\infty,
\end{equation}
for $1\leq p<\infty$ and 
\begin{equation}
  \label{eq:Yinfty}
  \|g\|_{Y_\infty} = \sup\limits_{\bu,\bxi\in\Rm^d}
    |\mathcal F h(\bu,\bxi)| <\infty.
\end{equation}
Finally, we define $Y$ as the subspace of 
tempered distributions in $\mathcal S'(\Rm^{2d})$ such that
\begin{equation}
  \label{eq:Y}
  \|g\|_{Y} = \sup\limits_{\bxi\in\Rm^d}\dint_{\Rm^d}
    |\mathcal F g(\bu,\bxi)|  d\bu <\infty.
\end{equation}

Morally (though this is inexact), the space $X_1$ corresponds to
scintillation functions that are integrable in one spatial variable
(bounded in the corresponding dual variable $\bv$) and bounded in
another spatial variable (integrable in the corresponding dual
variable $\bu$).  It is this boundedness that allows us to obtain the
result \eqref{eq:Jeta2} in the presence of small domains of
measurements. In contrast, $X_\infty$ corresponds to scintillation
functions that are integrable in both spatial variables (bounded in
$\bu$ and $\bv$), which allows us to get the result \eqref{eq:Jeta1}.

The above spaces are well-adapted to the estimation of the
scintillation operator $K_\eta$. More precisely, we have the following
result:
\begin{lemma}
  \label{lem:estimK}
  Assume that $\hat{R} \in L^1(\Rm^d) \cap L^\infty(\Rm^d)$. Then
  for $1\leq p\leq \infty$,
  \begin{itemize}
  \item[(i)] $K_\eta$ is bounded in $X_p$ and 
    \begin{equation}
      \label{eq:Ketaxp}
       \| \calK_\eta \|_{\calL(X_p)} \leq 4 \|\hat{R}\|_{L^1(\Rm^d)}.
    \end{equation}
   \item[(ii)] Let $\mu\in Y_p$ and $\nu\in Y$. Then
     \begin{equation}
        \label{estim:A2}
   \| \calK_\eta \, \mu \otimes \nu\|_{X_p} \leq 4 \,\eta^d \, 
   \|\hat{R}\|_{L^\infty(\Rm^d)} 
   \| \mu \|_{Y_p} \| \nu \|_{Y}.
     \end{equation}
  \end{itemize}
\end{lemma}
\begin{proof}
  With obvious notation, we recast
  $K_\eta=\sum_{\epsilon_i,\epsilon_j}\epsilon_i\epsilon_jK_\eta^{ij}$.
   Let  $h \in X_p$. Then we have
  \begin{eqnarray*}
  \calF \calK^{ij}_\eta\, h &=&
  \int_{\Rm^d} e^{i \bw\cdot (\frac{1}{2} \epsilon_i \bxi+\frac{1}{2} 
   \epsilon_j \bzeta)}\hat{R}(\bw) \calF h\left(\bu-\frac{\bw}{\eta},\bxi,\bv  
   +\frac{\bw}{\eta},\bzeta\right) d\bw,
  \end{eqnarray*} 
  so that using the H\"older inequality with $1=\frac{1}{p}+\frac{1}{p'}$,
  \begin{eqnarray*}
  &&\| \calK^{ij}_\eta\, h \|^p_{X_p}  \leq  \sup_{\bv,\bzeta \in \Rm^d} 
  \int_{\Rm^d}\sup_{\bxi  \in \Rm^d}  \left| \int_{\Rm^d} |\hat{R}(\bw) 
  \calF h\left(\bu-\frac{\bw}{\eta},\bxi,\bv+\frac{\bw}{\eta},\bzeta\right) 
  |d\bw\right|^p d\bu,\\
  &\leq &  \|\hat{R}\|^{\frac p{p'}}_{L^1(\Rm^d)} \sup_{\bv,\bzeta\in \Rm^d}
  \int_{\Rm^d}\sup_{\bxi \in \Rm^d} \int_{\Rm^d} |\hat{R}(\bw)| 
  \left|\calF h\left(\bu-\frac{\bw}{\eta},\bxi,\bv+\frac{\bw}{\eta},\bzeta 
   \right) \right|^p d\bw d\bu,\\
  &\leq & \|\hat{R}\|^p_{L^1(\Rm^d)}  \, \| h\|^p_{X_p}.
  \end{eqnarray*} 
  This proves (i). Let now $h:=\mu \otimes \nu$. Upon performing the change
  of variables $\bw \to \eta \bw$, we have
  \begin{eqnarray*}
  \calF \calK^{ij}_\eta\, \mu \otimes \nu &=&\eta^d
  \int_{\Rm^d} e^{i \eta \bw\cdot (\frac{1}{2} \epsilon_i \bxi+\frac{1}{2} 
  \epsilon_j \bzeta)}\hat{R}(\eta \bw) \calF \mu \otimes 
    \nu\left(\bu-\bw,\bxi,\bv+\bw,\bzeta\right) d\bw,
  \end{eqnarray*}
  so that 
  \begin{displaymath}
    \begin{array}{l}
  \| \calK^{ij}_\eta\, h \|^p_{X_p}  \\
   \leq   \eta^d\sup\limits_{\bv,\bzeta \in \Rm^d} 
  \dint_{\Rm^d} \sup\limits_{\bxi \in \Rm^d} \left| \dint_{\Rm^d} 
   |\hat{R}(\eta (\bw-\bv)) \calF \mu \otimes \nu 
  \left(\bv+\bu-\bw,\bxi,\bw,\bzeta\right) |d\bw\right|^p d\bu,\\
  \leq \eta^d\,  \|\hat{R}\|^p_{L^\infty(\Rm^d)} \,\| \nu\|^{\frac p{p'}}_{Y} 
  \sup\limits_{\bv,\bzeta  \in \Rm^d} \dint_{\Rm^d} 
   \sup\limits_{\bxi \in \Rm^d} 
  \dint_{\Rm^d} |\calF \mu(\bv+\bu-\bw,\bxi)|^p |\calF\nu 
  \left(\bw,\bzeta\right)  |d\bw d\bu,\\
  \leq   \eta^d\, \|\hat{R}\|^p_{L^\infty(\Rm^d)}  \, \| \mu\|^p_{Y_p} \,  
  \| \nu\|^p_{Y},
    \end{array}
  \end{displaymath}
  which concludes our proof.
\end{proof}

We need stability estimates for the kinetic equations. We start with
the first kinetic equation:
\begin{eqnarray} \label{eq:transport}
\frac{ \partial a}{\partial t} +\bp \cdot \nabla_\bx  a+R_0 
  \,a&=&\calQ a+S, \qquad a(0, \bx,\bp)=a_0(\bx,\bp), \\  \nonumber
\calQ a(t,\bx,\bp)&=&(2 \pi)^{-d}\int_{\Rm^d} \hat{R}(\bp-\bp')
a(t,\bx,\bp') d\bp',
\end{eqnarray}
with $R_0:=R(\bzero)$, $R \in L^1(\Rm^d)\cap L^\infty(\Rm^d)$ and
$\hat{R}$ non-negative. Then we have:
\begin{lemma} \label{wellposed1} 
  Assume that $a_0 \in Y_p$ and $S \in L^1((0,T),Y_p)$ for some $T>0$
  and $1\leq p \leq \infty$. Then (\ref{eq:transport}) admits a unique
  solution in $\calC^0([0,T],Y_p)$ such that
  \begin{equation} \label{estim:transpL1}
   \|a\|_{\calC^0([0,T],Y_p)} \leq \|a_0\|_{Y_p}+\|S\|_{L^1((0,T),Y_p)}.
  \end{equation}
  Let $S=0$ and let $a^0(t,\bx,\bp):=a_0(\bx-t\bp,\bp)
  e^{-R_0t}$ be the ballistic part of $a$.  Then, assuming that
  $\calF_\bk a_0 \in L^1(\Rm^{2d})$, we have the following estimate for
  all $t>0$:
  \begin{equation} \label{estim:transpL2}
  \|(a-a^0)(t,\cdot)\|_{Y} 
  \lesssim t^{1-d } \dint_{\Rm^d} \sup\limits_{\bv\in\Rm^d}
  |\mathcal F a_0(\bv,\bxi)| d\bxi
   \lesssim  t^{1-d }  \|\calF_\bk a_0\|_{L^1(\Rm^{2d})}.
  \end{equation}
\end{lemma} 
\begin{proof}
  The proof is a direct application of the integral formulation of
  (\ref{eq:transport}),
  $$
  a(t)=e^{-R_0 t} \calG_t a_0+\int_0^t e^{-R_0 (t-s)}
  \calG_{t-s}\calQ (a(s)+S(s))ds,
  $$
  where $\calG_t$ is the free transport semigroup given by
  $$
  \calG_t a(\bx,\bp):=a(\bx-t\bp,\bp).
  $$
  The operators $\calQ$ and $\calG_t$ are both continuous in $Y_p$.
  Indeed, for $\varphi \in Y_p$, we have:
  \begin{eqnarray*}
\calF \calG_t \varphi &=& \calF \varphi(\bu,\bxi+t\bu),\\
\calF \calQ \varphi   &=& R(\bxi) \calF\varphi(\bu,\bxi),
  \end{eqnarray*}
  so that
  \begin{equation} \label{eq:bdtr2}
    \begin{array}{rcl}
  \|\calG_t \varphi \|_{Y_p}&\leq& \|\varphi \|_{Y_p},\\
  \| \calQ \varphi  \|_{Y_p} &\leq & \|R\|_{L^\infty(\Rm^d)}  
  \|\varphi \|_{Y_p}.
    \end{array}
  \end{equation}
  Standard fixed point techniques then provide existence and
  uniqueness results for (\ref{eq:transport}). When $S=0$, estimate
  (\ref{estim:transpL1}) follows from the maximum principle and the
  observation that $\|a_0\|_{Y_p}$ is a majorizing solution to
  (\ref{eq:transport}). When $a_0=0$, (\ref{estim:transpL1}) is an
  application of the Gronwall lemma.
  
  For $S=0$, we have the following Neumann series expansion in terms
  of multiple scattering:
  $$
  a^n(t)=\int_0^t e^{-R_0(t-s)}\calG_{t-s} \calQ a^{n-1}(s) ds,
  $$
  with the ballistic part $a^0(t,\bx,\bp):=e^{-R_0 t}
  a_0(\bx-t\bp,\bp)$. By induction, we find the following expression
  for the Fourier transform of $a^n$:
  \begin{eqnarray*}
  \calF a^n(t,\bu,\bk)&=& e^{-R_0 t}\int_0^t\int_0^{s_1}\cdots\int_0^{s_{n-1}} 
  R(\bk+(t-s_1) \bu )\cdots \\&&\qquad\qquad R(\bk+(s_{n-1}-s_n)\bu)
   \calF a_0(\bu,\bk+t\bu)ds_1 \cdots ds_n.
  \end{eqnarray*}  
  The change of variable $\bk+t \bu \to \bxi $ yields
  \begin{eqnarray*}
  \|a^n(t,\cdot)\|_{Y} &\leq &\frac{e^{-R_0 t}}{n!\, 
    t^{d-n}}\|R\|^{n}_{L^\infty(\Rm^d)} \int_{\Rm^d} \sup_{\bv \in \Rm^d} 
  |\calF a_0(\bv,\bxi)| d\bxi, \\
   &\leq &\frac{e^{-R_0 t}}{n!\, 
    t^{d-n}}\|R\|^{n}_{L^\infty(\Rm^d)} 
   \| \calF_\bk a_0\|_{L^1(\Rm^{2d})}.
  \end{eqnarray*} 
  Summing over $n\geq1$ gives the result.
\end{proof}

The last lemma deals with the fourth-order transport equation
\eqref{eq:kinfour}:
\begin{lemma} \label{wellposed2} 
  Assume $a_0 \in X_p$ and $S \in L^1((0,T),X_p)$, for $T>0$ and
  $1 \leq p \leq \infty$. Then, the above system admits a unique
  solution in $\calC^0([0,T], X_p)$ such that:
 \begin{equation} \label{estim:transpE}
  \|a\|_{\calC^0([0,T],X_p)} \leq \|a_0\|_{X_p}+\|S\|_{L^1((0,T),X_p)}.
  \end{equation}
\end{lemma}
\begin{proof} 
  The result stems from the integral formulation of \eqref{eq:kinfour}
  given by
  $$
  a(t)=e^{-2 R_0 t} \calG^2_t a_0+\int_0^t e^{-2R_0 (t-s)} 
  \calG^2_{t-s}[(\calQ_2+\calK_\eta) a+S](s)ds,
  $$
  where $ \calG^2_t$ is the semigroup defined as
  $$
  \calG^2_t a(\bx,\bp,\by,\bq):=a(\bx-t\bp,\bp,\by-t\bq,\bq).
  $$
  From Lemma \ref{lem:estimK}, we know $\calK_\eta$ is continuous
  in $X_p$, and so are $\calG^2_t$ and $\calQ_2$ since
  \begin{eqnarray*}
\calF \calG^2_t \varphi &=& \calF \varphi(\bu,\bxi+t\bu,\bv,\bzeta+t\bv),\\
\calF \calQ_2 \varphi &=&  (R(\bxi)+R(\bzeta)) \calF  
\varphi(\bu,\bxi,\bv,\bzeta),
  \end{eqnarray*}
  for $\varphi \in X_p$. Existence and uniqueness follow as before
  from standard fixed point theorems while estimate
  (\ref{estim:transpE}) stems from separate applications of the
  maximum principle and the Gronwall lemma.
\end{proof}

\section{Estimates for the scintillation}
\label{sec:estimates}

We are now ready to prove Theorem \ref{thm:1} and Corollary
\ref{cor:1}.
\begin{proofof} [Theorem \ref{thm:1}].
  According to Lemma \ref{wellposed2}, the fourth-order transport
  equation (\ref{eq:J}) is stable in $X_p$, so that we have the
  following estimate, uniformly on $[0,T]$,
  \begin{equation}\label{eq:estimJeta}
    \|J_\eta(t)\|_{X_p} \lesssim 
   \int_0^t \| \calK_\eta a_\eta \otimes a_\eta \, (s)\|_{X_p} ds.
  \end{equation}
  Provided that $a_\eta$ belongs to $Y \cap Y_p$, then $\calK_\eta
  a_\eta \otimes a_\eta $ is small in $X_p$. Indeed, item (ii) of
  Lemma \ref{lem:estimK} yields for $s \in [0,T]$ and $1 \leq p \leq
  \infty$ that:
  $$
  \| \calK_\eta a_\eta \otimes a_\eta \, (s)\|_{X_p} \leq 4 \eta^d
  \|\hat{R} \|_{L^\infty(\Rm^d)} \|a_\eta(s)\|_{Y} \,
  \|a_\eta(s)\|_{Y_p}.
  $$
  First, we control the $Y$ norm by the $Y_1$ norm since
  $Y_1\subset Y$. Lemma \ref{wellposed1} shows that the radiative
  transfer equation (\ref{eq:kinetic}) is stable in $Y_r$, for $1\leq
  r \leq \infty$, so that we just need to estimate the initial
  condition $a_{\eta0}(\bx,\bp):=a_\eta(0,\bx,\bp)$ in these $Y_r$
  norms. Denoting by $\calF_\bx a_{\eta0}(\bu,\bp)$ the Fourier
  transform of $a_{\eta0}$ with respect to the spatial variable $\bx$
  only, we obtain,
  \begin{eqnarray*}
     \| a_{\eta0}\|_{Y_1} &\leq& \int_{\Rm^{2d}} \left| 
         \calF_\bx a_{\eta0}(\bu,\bp) \right|d\bp d\bu,\\
        \| a_{\eta0}\|_{Y_\infty} &\leq& \sup_{\bu \in \Rm^{d}} \int_{\Rm^d} | 
     \calF_\bx a_{\eta0} (\bu,\bp) |d\bp,
  \end{eqnarray*}
  so that the assumption of the theorem gives
  $$
  \| a_{\eta0}\|_{Y_1} \leq C \eta^{-d\alpha}.
  $$
  Moreover, defining $\psi_{\eta0}(\cdot):=\psi_\eta(\cdot,0)$, we
  have the relation
  $$
  \calF_\bx a_{\eta0}(\bu,\bp)=\frac{1}{\eta^d} \calF
  \psi_{\eta0}\left( \frac{\bp}{\eta}+\frac{\bu}{2}\right)
  \overline{\calF \psi_{\eta0}}\left(
    \frac{\bp}{\eta}-\frac{\bu}{2}\right),
  $$
  from which it follows, using the Cauchy-Schwarz inequality, that
  $$
  \| a_{\eta0}\|_{Y_\infty} \leq \| \calF \psi_{\eta0}
  \|^2_{L^2(\Rm^d)} \leq C,
  $$
  where $C$ is independent of $\eta$. We have thus obtained that
  for all $s\in[0,T]$,
  \begin{eqnarray*}
    \| \calK_\eta a_\eta \otimes a_\eta \, (s)\|_{X_\infty} 
    &\leq& C \eta^{d(1-\alpha)},\\
   \| \calK_\eta a_\eta \otimes a_\eta \, (s)\|_{X_1} &\leq& 
   C \eta^{d(1-2\alpha)},
  \end{eqnarray*}
  which yields by interpolation, for $1\leq p \leq \infty$,
  \begin{equation} \label{estimJ1}
    \| \calK_\eta a_\eta \otimes a_\eta \, (s)\|_{X_p} \leq 
    C \eta^{d\left(1-(1+\frac{1}{p})\alpha \right)}.
  \end{equation}
  This induces a first estimate for $J_\eta$, which is not optimal for
  initial conditions with small support when $\alpha-\beta>0$. The
  stability of the transport equation \eqref{eq:kinetic} in $Y_p$ is
  not sufficient to deal with such irregular initial conditions.
  Rather, we need to separate the ballistic part from the scattering
  part in the kinetic equation to obtain sharper estimates and thus
  introduce:
  $$
  a_\eta(t,\bx,\bp):=a_\eta^{0}(t,\bx,\bp)+a_\eta^{\textrm{s}}
  (t,\bx,\bp),
  $$
  where $a_\eta^{0}(t,\bx,\bp)=e^{-R_0 t}
  a_{\eta0}(\bx-t\bp,\bp)$ is the ballistic part and
  $a_\eta^{\textrm{s}}$ satisfies
  $$
  \frac{\partial a^\textrm{s}_\eta}{\partial t} +\bp \cdot
  \nabla_\bx a^\textrm{s}_\eta+R_0 \,a^\textrm{s}_\eta=\calQ
  a^\textrm{s}_\eta+\calQ a^0_\eta , \qquad
  a^\textrm{s}_\eta(t=0, \bx,\bp)=0.
  $$
  Since the Fourier transform of $a^0_\eta$ is given by $e^{-R_0 t}
  \calF a_{\eta0}(\bu,\bk+t\bu)$, its $Y$ norm can be estimated
  for $t \in(0,T]$ as:
  \begin{eqnarray*}
  \| a_\eta^0(t)\|_{Y} &\leq& \sup_{\bk \in \Rm^d}\int_{\Rm^{d}} \left| 
  \calF a_{\eta0}(\bu,\bk+t\bu) \right| d\bu\,\,
  \leq \,\,\frac{1}{t^d} \sup_{\bk \in \Rm^d}\int_{\Rm^{d}} \left| 
  \calF a_{\eta0}(t^{-d} (\bk-\bz),\bz) \right| d\bz\\
  &\leq& \frac{1}{t^d} \int_{\Rm^{2d}} \left| \calF_\bk a_{\eta0}(\bx,\bz) 
  \right| d\bz d\bx\,\, \leq \,\,  \frac{C}{t^d} \eta^{-d\beta}.
  \end{eqnarray*}
  Now, Lemma \ref{wellposed1} and estimate (\ref{estim:transpL2})
  imply that:
  $$
  \| a_\eta^s(t)\|_{Y} \leq \frac{C}{t^{d-1}} \eta^{-d\beta},
  $$
  so that the time singularity of $a^\textrm{s}_\eta$ is weaker
  than that of $a_\eta^{0}$. Thus, for $1\leq p\leq \infty$, 
  $$
  \| \calK_\eta a_\eta \otimes a_\eta \, (s)\|_{X_p} \lesssim
  \eta^{d\left(1-\beta-\frac{1}{p}\alpha \right)}\, (s^{-d}+s^{1-d})
  \lesssim \eta^{d\left(1-\beta-\frac{1}{p}\alpha \right)} s^{-d}.
  $$
  For short times, we then use estimate \eqref{estimJ1} since it is
  independent of $s$ and for longer times, we use the above estimate.
  We thus write:
  $$
  \| \calK_\eta a_\eta \otimes a_\eta \, (s)\|_{X_p}=\un\left(s
    \leq t_0(\eta)\right) \| \calK_\eta a_\eta \otimes a_\eta \,
  (s)\|_{X_p}+\un\left(s > t_0(\eta)\right) \| \calK_\eta a_\eta
  \otimes a_\eta \, (s)\|_{X_p},
  $$
  so that, for $t \in [0,T]$, we have
  $$
  \|J_\eta(t)\|_{X_p} \leq C \, t_0(\eta)\,
  \eta^{d\left(1-(1+\frac{1}{p})\alpha \right)} + C
  t_0^{1-d}(\eta) 
  \eta^{d\left(1-\beta-\frac{1}{p}\alpha \right)}.
  $$
  Setting $t_0(\eta)=\eta^{\alpha-\beta}$ when $\alpha>\beta$ above
  and using \eqref{estimJ1} and \eqref{eq:estimJeta}, we find, for $t
  \in [0,T]$, that
  $$
  \|J_\eta(t)\|_{X_p} \leq C
  \eta^{d\left(1-(1+\frac{1}{p})\alpha
    \right)+(\alpha-\beta) \vee 0}.
  $$
  We conclude by using the Parseval-Plancherel equality which
  yields, for $t \in [0,T]$,
  \begin{eqnarray*}
  &&\left| \big \langle J_\eta, \varphi_{\eta,s_1,s_2} \otimes 
   \varphi_{\eta,s_1,s_2} \big \rangle \right| = (2 \pi )^{-d} 
 \left|\big \langle \calF J_\eta, \calF \varphi_{\eta,s_1,s_2} \otimes 
  \varphi_{\eta,s_1,s_2} \big \rangle \right|,\\
  &\leq &(2 \pi )^{-d} \, \|J_\eta\|_{X_p} \, \|  \calF  
  \varphi_{\eta,s_1,s_2} \|_{L^1(\Rm^{2d})} \,  \left\|  \int_{\Rm^d} |\calF  
   \varphi_{\eta,s_1,s_2}(\cdot,\bp)|d\bp \right\|_{L^{p'}(\Rm^d)},
  \end{eqnarray*}
  with $1=\frac{1}{p}+\frac{1}{p'}$. It remains to verify the scaling
  properties:
  \begin{eqnarray*}
     \|  \calF  \varphi_{\eta,s_1,s_2} \|_{L^1(\Rm^{2d})}
    &=&\frac{1}{\eta^{d(s_1+s_2)}} \|  \calF  \varphi\|_{L^1(\Rm^{2d})},\\
   \left\|  \int_{\Rm^d} |\calF  \varphi_{\eta,s_1,s_2}(\cdot,\bp)|d\bp 
    \right\|_{L^{p'}(\Rm^d)}&=&\frac{1}{\eta^{d\left(s_1(1-\frac{1}{p})
     +s_2\right)}} \left\|  \int_{\Rm^d} |\calF  \varphi(\cdot,\bp)|d\bp 
   \right\|_{L^{p'}(\Rm^d)}.
  \end{eqnarray*}
  We conclude the proof of the theorem by choosing $p=\infty$ or $p=1$
  in the above estimates.
\end{proofof}
\begin{proofof}[Corollary \ref{cor:1}].
  We simply need to estimate $\calF_\bx a_{0\eta}$ and $\calF_\bk
  a_{0\eta}$ in $L^1(\Rm^d)$. Since
  \begin{eqnarray*}
\calF_\bx a_{\eta0}(\bu,\bp)&=&\frac{1}{\eta^d} \calF \psi_{\eta0}\left( \frac{\bp}{\eta}+\frac{\bu}{2}\right)  \overline{\calF \psi_{\eta0}}\left( \frac{\bp}{\eta}-\frac{\bu}{2}\right),\\
\calF_\bk a_{\eta0}(\bx,\bk)&=&\psi_{\eta0}\left( \bx+\frac{\eta}{2}\bk\right)  \overline{\psi_{\eta0}}\left(\bx-\frac{\eta}{2}\bk\right),
  \end{eqnarray*}
  it follows that:
  \begin{eqnarray*}
    &&\int_{\Rm^{2d}} |\calF_\bx a_{\eta0}(\bu,\bp)| d\bu d\bp
    =\frac{1}{\eta^d} \int_{\Rm^{2d}} \left|\calF \psi_{\eta0}
    \left( \frac{\bp}{\eta}+\frac{\bu}{2}\right)  
     \overline{\calF \psi_{\eta0}}\left( \frac{\bp}{\eta}
    -\frac{\bu}{2}\right)\right|  d\bu d\bp, \\
       &&=\int_{\Rm^{2d}}  \left|\calF \psi_{\eta0}\left(\bu\right) 
     \overline{\calF \psi_{\eta0}}\left(\bp\right)\right|d\bu d\bp\,\,
   =\,\,\|\calF \psi_{\eta0}\|^2_{L^1(\Rm^d)} \,\, \leq \,\, C \eta^{-d\alpha},
  \\
     &&\int_{\Rm^{2d}} |\calF_\bk a_{\eta0}(\bx,\bp)| d\bx d\bp\leq 
   \eta^{-d}\|\psi_{\eta0}\|^2_{L^1(\Rm^d)} \,\,\leq\,\, C\eta^{-d(1-\alpha)}.
  \end{eqnarray*}
  It suffices to set $\beta=1-\alpha$ in Theorem \ref{thm:1} to
  conclude the proof of the corollary.
\end{proofof}

\section{Convergence of the scintillation}
\label{sec:conv}

We now prove the announced convergence result. We first observe that
the existence results obtained in Lemmas \ref{wellposed1} and
\ref{wellposed2} hold when the spaces $Y_p$ and $X_p$ are replaced by
the spaces of bounded measures $\mathcal M(\Rm^{2d})$ and $\mathcal
M(\Rm^{4d})$, respectively or by the spaces of continuous functions
$\mathcal C^0(\Rm^{2d})$ and $\mathcal C^0(\Rm^{4d})$, respectively.
We recall that $d\geq2$ here.

\begin{proofof}[Theorem \ref{thm:2}].
  The scintillation function satisfies the following transport
  equation in integral form
  \begin{equation}
    \label{eq:intJ}
     J_\eta(t) =\dint_0^t e^{-2R_0(t-s)}\mG^2_{t-s}
    (\mathcal Q_2+K_\eta)J_\eta(s)ds
    + \dint_0^t e^{-2R_0(t-s)} \mG^2_{t-s}  K_\eta a\otimes a (s)ds.
  \end{equation}
  We recast this, with obvious notation, as
  \begin{displaymath}
  J_\eta= T_{2\eta} J_\eta + J_\eta^0 ,\qquad J_\eta = \dsum_{k=0}^\infty
    T_{2\eta}^k  J_\eta^0.
  \end{displaymath}
  We denote by $T_2$ the formal limit operator of $T_{2\eta}$ defined
  as
  \begin{equation}
    \label{eq:T2}
    T_2f(t)=\dint_0^t e^{-2R_0(t-s)}\mG^2_{t-s}\mathcal Q_2f(s)ds.
  \end{equation}

  \noindent{\bf The source contribution.}
  We verify that 
  \begin{equation}
    \label{eq:errorKeta}
    \|K_\eta a\otimes a (s) - K_\eta a^0\otimes a^0(s)\|_{X_\infty}
     \lesssim \dfrac{\eta^d}{s^{d-1}}\wedge 1 ,
  \end{equation}
  where the ballistic part is given by 
  \begin{displaymath}
  a^0(t,\bx,\bk) = e^{-R_0t} \delta(\bx-t\bk)f(\bk)=e^{-R_0t}\dfrac{1}{t^d}
   \delta(\bk-\frac \bx t) f(\frac \bx t).
  \end{displaymath}
  Indeed, we know from Lemma \ref{lem:estimK}
  that 
  \begin{displaymath}
    \|K_\eta (a-a^0)\otimes a (s)\|_{X_\infty} \lesssim \eta^d
     \|a-a^0\|_Y \|a\|_{Y_\infty},
  \end{displaymath}
  and from \eqref{estim:transpL2} in Lemma \ref{wellposed1} that 
  \begin{displaymath}
     \|a-a^0\|_Y \lesssim t^{1-d } \dint_{\Rm^d} 
     \sup\limits_{\bv\in\Rm^d}
     |\mathcal F a_0(\bv,\bxi)| d\bxi \lesssim t^{1-d } 
         \dint_{\Rm^d}|\hat f(\bxi)|d\bxi.
  \end{displaymath}
  That $\|a\|_{Y_\infty}$ is bounded comes from the stability of the
  transport equation in $Y_\infty$ established in Lemma
  \ref{wellposed1}.  The term $K_\eta a^0\otimes(a-a^0)$ is treated
  similarly.  Let us define
  \begin{equation}
  \label{eq:Jeta00}
   \begin{array}{rcl}
  J_\eta^{00}(t) &=& \dint_0^t e^{-2R_0(t-s)} \mG^2_{t-s}  
         K_\eta a^0\otimes a^0 (s)ds,\\
   &=& e^{-2R_0t} \dint_0^t \mG^2_{t-s} K_\eta
        \mG^2_s ( a_0\otimes a_0 )ds.
   \end{array}
  \end{equation}
  We find that
  \begin{equation}
    \label{eq:errorJ}
    \|J^0_\eta(t) - J^{00}_\eta(t)\|_{X_\infty}
    \lesssim \eta^{\frac d{d-1}} \ll \eta.
  \end{equation}
  Indeed, we deduce from \eqref{eq:errorKeta} and the stability of
  $\mG^2_{t}$ in $\mathcal L(X_\infty)$ that
  \begin{displaymath}
    \begin{array}{l}
        \|J^0_\eta(t) - J^{00}_\eta(t)\|_{X_\infty}
     \lesssim \dint_0^t (\dfrac{\eta^d}{s^{d-1}}\wedge 1) ds
    \lesssim t_0+\eta^{d}t_0^{2-d} \lesssim \eta^{\frac d{d-1}},
    \end{array}
  \end{displaymath}
  for $t_0=\eta^{\frac{d}{d-1}}$. Up to a smaller-order error term in
  the space of distributions, we may thus replace $J^0_\eta$ by
  $J^{00}_\eta$ in the sequel since the transport equation
  \eqref{eq:intJ} is stable in $X_\infty$. Now, calculations with
  $K_\eta$ replaced by $K_{\eta}^{-1,-1}$ show that
  \begin{displaymath}
   \begin{array}{rcll}
   e^{2R_0t}J_{\eta,11}^{00}(t)
   &=& \dint_0^t \dint_{\Rm^d} \hat R(\bu)&
   e^{i\frac \bu\eta\cdot [(\bx-(t-s)\bp)-(\by-(t-s)\bq)]}
   \delta(\bx-t\bp+s\frac \bu2)  \\
   &&&  \delta(\by-t\bq+s\frac \bu2)
   f(\bp-\frac \bu2) f(\bq-\frac \bu2) d\bu ds \\
    &=& \dint_0^t \dint_{\Rm^d} \hat R(\bu)&
   e^{i\frac {\bu s}\eta\cdot (\bp-\bq)}
   \delta(\bx-t\bp+s\frac \bu2)  \\
   &&&  \delta(\by-t\bq+s\frac \bu2)
   f(\bp-\frac \bu2) f(\bq-\frac \bu2) d\bu ds ,
   \end{array}
  \end{displaymath}
  \begin{displaymath}
   \begin{array}{rcll}
    e^{2R_0t}J_{\eta,11}^{00}(t) 
   &=& \eta \dint_0^{\frac t\eta} \dint_{\Rm^d} \hat R(\bu)&
   e^{i s\bu\cdot (\bp-\bq)}
   \delta(\bx-t\bp+\eta s\frac \bu2)  \\
   &&&  \delta(\by-t\bq+\eta s\frac \bu2)
   f(\bp-\frac \bu2) f(\bq-\frac \bu2) d\bu ds.
   \end{array}
  \end{displaymath}
  Upon sending $\eta\to0$, we find in the limit that 
  \begin{displaymath}
     \begin{array}{l}
  \lim\limits_{\eta\to0} \dfrac{J^{00}_{\eta,11}(t)}{\eta}
  = J^{0}_{11}(t) = \\ e^{-2R_0t}\delta(\bx-t\bp)\delta(\by-t\bq)
     \pi \dint_{\Rm^d} \hat R(\bu) \delta\big(\bu\cdot(\bp-\bq)\big) 
   f(\bp-\frac \bu2) f(\bq-\frac \bu2) d\bu\\
  +i e^{-2R_0t}\delta(\bx-t\bp)\delta(\by-t\bq)
    \textrm{ p.v.}\dint_{\Rm^d} \hat R(\bu)\, \frac{1}{\bu\cdot(\bp-\bq)} 
   f(\bp-\frac \bu2) f(\bq-\frac \bu2) d\bu,
    \end{array} 
  \end{displaymath}
  in the space of bounded measures $\mathcal M(\Rm^{4d})$. After
  accounting for all four terms in the definition of $K_\eta$ and
  using the fact that $\hat R(\bu)=\hat R(-\bu)$, we find that the
  limit of $\eta^{-1}J^0_\eta(t)$ is given by:
  \begin{displaymath}
    \begin{array}{ll}
    J^0(t) = &e^{-2R_0t}
  \delta(\bx-t\bp)\delta(\by-t\bq) \times \\&2\pi\dint_{\Rm^d}
   \hat R(\bu) \delta\big(\bu\cdot(\bp-\bq)\big) 
    f(\bp-\dfrac{\bu}2)\Big( f(\bq-\dfrac{\bu}2)-f(\bq+\dfrac{\bu}2)\Big)d\bu.
   \end{array}  
  \end{displaymath}
  This gives us the source term \eqref{eq:J0} in the transport
  equation \eqref{eq:J}.

  \medskip
  \noindent{\bf Kinetic equation for the scintillation.}
  We have shown that $\eta^{-1}J^0_\eta$ converged to $J^0$. It
  remains to obtain convergence of the whole sequence
  $\eta^{-1}J_\eta$. Let $\phi(t,\bx,\bp,\by,\bq)$ be a a smooth
  function on $[0,T]\times \Rm^{4d}$.  Then we have by integration on
  the latter space that
  \begin{equation}
  \label{eq:weakint}
  (J_\eta,\phi) = (T_{2\eta}J_\eta,\phi) + (J^0_\eta,\phi),
  \end{equation}
  and equivalently that
  \begin{equation}
  \label{eq:weakadj}
  (J_\eta,\phi) = (J_\eta,T_{2\eta}^*\phi) + (J^0_\eta,\phi),
  \end{equation}
  with
  \begin{equation}
  \label{eq:T2adj}
  T_{2\eta}^* \phi (s) = \dint_s^T e^{-2R_0(t-s)}
   (\mathcal Q_2^* + K_\eta^*) \mathcal G^{2*}_{t-s} \phi (t) dt.
  \end{equation}
  
  We have shown that the difference between the source terms
  $\eta^{-1}J^0_\eta$ and $J^0$ converges to $0$ as a distribution and
  has a negligible effect on $\eta^{-1}J_\eta$. So we can replace the
  initial condition for the error term by $J^0$ and look at the
  problem
  \begin{displaymath}
  \tilde J_\eta = T_{2\eta} \tilde J_\eta + J^0,
  \end{displaymath}
  where $\tilde J_\eta$ is now of order $O(1)$. We observe that
  \begin{displaymath}
  J^0(t) = e^{-2R_0t}\delta(\bx-t\bp)\delta(\by-t\bq) H(\bp,\bq),
  \end{displaymath}
  where $H(\bp,\bq)$ is a smooth function.
  
  Let now $J^1_\eta=\tilde J_\eta-J^0$ be the solution of
  \begin{displaymath}
  J^1_\eta = T_{2\eta} J^1_\eta + T_{2\eta} J^0.
  \end{displaymath}
  We recall that
  \begin{displaymath}
  T_{2\eta} J(t) = 
  \dint_0^t e^{-2R_0(t-s)}\mG^2_{t-s}(\mathcal Q_2+K_\eta)J(s)ds,
  \end{displaymath}
  so that
  \begin{displaymath}
  T_{2\eta} J^0 = T_2 J^0 + J^2_\eta, 
  \end{displaymath}
  where $J^2_\eta$ is given by a bounded operator in $\mathcal
  M(\Rm^{4d})$ applied to $K_\eta J^0$. The latter is given by
  \begin{displaymath}
  \delta(\bx-t\bp)\delta(\by-t\bq)
    e^{-2R_0t}\dint_{\Rm^d} \hat R(\bu) e^{i\frac{\bp-\bq}{t\eta}\cdot \bu}
    H(\bp-\frac \bu2) H(\bq-\frac \bu2)d\bu,
  \end{displaymath}
  plus similar contributions. Because $H$ is a smooth function, this
  term converges to $0$ in $\mathcal M(\Rm^{4d})$ as $\eta\to0$. This
  shows that $J^2_\eta$ converges to $0$ as $\eta\to0$.
  
  The other contribution, $T_2 J^0$, involves a bounded operator
  applied to $\mathcal Q_2 J^0$, which is equal to
  \begin{equation} \label{eq:Q2J0}
  \mathcal Q_2 J^0(t) (\bx,\bp,\by,\bq)= e^{-2R_0t} \hat R(\bp-\frac{\bx}t)
   \hat R(\bq-\frac{\by}t) 
  \dfrac{1}{t^{2d}} H(\frac{\bx}t,\frac \by t).
  \end{equation}
  For $f$, whence $H$, and $R$ sufficiently smooth, the above function
  is bounded in $\mathcal C^0(\Rm^{4d})$. The function is not bounded
  uniformly in time, however, and we split the contribution $J^0(t)$
  into $J^0_\delta(t)=J^0\chi_{(0,\delta)}(t)$ and $J^0\chi_{(\delta,
    T)}(t)$, which we still denote by $J^0(t)$. The source term
  $T_{2}J^0_\delta$ generates a small contribution, which goes to $0$
  as $\delta$ goes to $0$ in the sense of distributions since the term
  in \eqref{eq:Q2J0} is bounded in e.g. $L^1(\Rm^{4d})$ uniformly in
  time so that after time integration in \eqref{eq:T2}, the
  contribution is bounded by $O(\delta)\to0$. The remaining
  contribution is bounded in the uniform norm uniformly in time with
  bound inversely proportional to $\delta^{2d}$.

  We now have a problem of the form
  \begin{displaymath}
  J^1_\eta = T_{2\eta} J^1_\eta + T_2 J^0,
  \end{displaymath}
  where $T_2 J^0$ is uniformly bounded in the uniform norm by
  $O(\delta^{-2d})$. Weakly, this means that 
  \begin{displaymath}
  (J^1_\eta,\phi) = (J^1_\eta, T_{2\eta}^* \phi) + (T_2 J^0,\phi),
  \end{displaymath}
  where $\phi(t,\bx,\bp,\by,\bq)$ is a smooth function.  The solution
  $J^1_\eta$ is bounded in $\mathcal C^0(\Rm^{4d})$ uniformly in $\eta$
  by stability of the fourth-order transport equation in the uniform
  norm.  There is therefore a subsequence that converges weak $*$ in
  $L^\infty(\Rm^{4d})$ to a limit $J^1\in L^\infty(\Rm^{4d})$.

  Let us decompose $T_{2\eta}$ as:
  \begin{displaymath}
  T_{2\eta} J(t) = 
  \dint_0^t e^{-2R_0(t-s)}\mG^2_{t-s}(\mathcal Q_2+K_\eta)J(s)ds
  = T_2 J(t) + S_{2\eta} J(t),
  \end{displaymath}
  where
  \begin{displaymath}
  S_{2\eta} J(t) = 
  \dint_0^t e^{-2R_0(t-s)}\mG^2_{t-s} K_\eta J (s)ds.
  \end{displaymath}
  We choose $\phi$ sufficiently smooth so that $S_{2\eta}^*\phi$ goes
  to $0$ strongly in $L^1(\Rm^{4d})$.  As a consequence,
  $(J^1_\eta,S_{2\eta}^*\phi)$ goes to $0$ with $\eta$ so that, in the
  limit, we have
  \begin{displaymath}
   (J^1,\phi) = (J^1,T_2^*\phi) + (T_2J^0,\phi).   
  \end{displaymath}
  The above convergence to the limiting transport equation holds for
  every cut-off $\delta$.  Thus, by stability of the limiting
  transport equation, we can remove the cut-off in $\delta$ and obtain
  that
  \begin{displaymath}
  J^1 = T_2 J^1 + T_2 J^0,
  \end{displaymath}
  weakly in the space of distributions. The above integral equation
  admits a unique solution, which shows that the whole sequence
  $\eta^{-1}J_\eta$ converges to $J$ solution of:
  \begin{displaymath}
  J = T_2 J + J^0.
  \end{displaymath}
  This completes the proof of Theorem \ref{thm:2}.
\end{proofof}

\section*{Acknowledgment}

This work was supported in part by DARPA-ONR Grant N00014-04-1-0224
and NSF Grant DMS-0239097.


\end{document}